# Smart Context-aware Rejuvenation of Engagement on Urban Ambient Augmented Things

Rossi Kamal, Choong Seon Hong, Kyung Hee University, South Korea

## ABSTRACT

The concern over global urbanization trend imposes smart-city as enabling information and communication technology (ICT) to improve urban governance. However, the light trance on better living space is stimulated by socio-economic impact of escalated senior generation. Hence, ambient assisted living (AAL) emerges for the autonomous provisioning of pervasive things or objects from relevant perturbation for advanced scientific instrumentation. Meanwhile, citizens are observed in being transfixed by lively stimuli of monotonous urban events with the advent of virtual reality or augmented things. Thus, due to the involvement of situation-awareness or contextualization, engagement/participation information as a utility promises to improve urban experience. However, it is complex to grapple meaningful concepts due to personalization obstacles, such as citizen psychology, information gap, service-visualization. Moreover, recommended practices deficit adaptation to monochromatic choice, disparate impairments, mobility and annotation-richness in urban space. Hence, rejuvenation of engagement relates to monitoring and quantification of 'service-consumption and graceful degradation' of experience. However, paramount challenges are imposed on this stipulation, such as, un-observability, independence and composite relationship of contexts. Therefore, a parametric Bayesian based model is envisioned to address observability and scalability of contexts and its conjugal relationship with engagement. Last but not the least, systematic framework is demonstrated, which pinpoints key goals of context-aware engagement from participants' opinions, usages and feed-backs.

## INTRODUCTION

Cities are reminiscent of verdict, culture and finance since primeval ages. Hence, the global urbanization trend is striving policy makers towards sustainable city, with improved governance and quality of life. Consequently, Smart-City concept emerges as an instance of Information and Communication Technology, which undertakes networked information for operational management of an urban-life. However, information is regarded as key player for multidimensional challenges, such as sharing, transfer and analysis of citizen knowledge-base. Therefore, Smart-City[1]-[3] is resembled by services that use, amongst others, Smart-devices, intelligent vehicles, and enable accumulation of monitoring information to react autonomously in real time, with the least possible human intervention.

However, social -economic impact of aging generation is striving policy makers for scientific instrumentation in a smarter urban life. Especially, they seem perturbed from the statistical predictions from UN and EC, that that aging generations exceed utilitarian for the first time in world around 2050 and with nearly 44% rise of seniors between 1995 and 2025 in only UK. However, citizens crave for soulful independent life with the trade-off for grant amount of his lifelong income. Such factors demand ambient technology with pervasive information, operation, management, and adequate permissibility for urban people. In this context, AAL is introduced by European Commission, synchronized with FP7 research area to promote slogans, such as, Aging Well, Prolonging Independent Living. This is given a new dimension with significance of recently launched program entitled Horizon 2020. AAL is meant to use ICT technology to sense information from objects, or things for autonomous provisioning of well-being for an independent life. This is a follow up of activity monitoring of 80s, which consequently expands through recent penetration of Smart-device and wearable devices. Eventually, scientific endeavors evolve from closed-door set-ups to miniature- scaled deployment with similar slogans, such as 'E-Health', 'Smart-Health', 'Connected Health', etc.

Citizens are so habituated with ambient technology, that they seldom bother about its acceptability. Eventually, they recently are overwhelmed by augmented reality in the nub of killer Smart city apps. The actual reason is beyond contemporary urban space, where they are looking for inner meaning of a virtual entertainment being exhausted with typical living. Virtual reality is the experience of in-depth perception, which inspires citizens in coming out of monotonous life and to be left alone in another pleasant virtual space. Stepping into this virtual world, citizen gets excited, which, eventually, has profound effect on him. Thus, citizen is transfixed by conventional urban events with the glimpse of augmented experience. Therefore, the virtual presence has the promise of apprehending lively stimuli to conventional urban space.

In this context, recently operators are shifting to 5G-driven solutions due to the changed networked environment for meeting the demand of immersive multimedia contents. As a follow up of it, a common trend is to connect every things or objects or humans in the form of Machine-to-Machine Communication or Machine-Type Communication. However, the increasing craving for personalized services are striving service providers in investigating on how , when and where citizens are connected to ambient Internet-of-Things. Thus, context aware revolution of ambient augmented things is expected to result in a Smarter urban life.

However, citizen engagement on ambient things is obligatory due to their multi-modal contexts, especially behavior and activity. Things or objects embedded on sensors, resemble surrounding contexts, including unusual or sudden change on psychological or physiological factors. Thus, adequate inference of engagement information govern caregivers' optimized decision making. Engagement seldom becomes intricate in small scale deployment, such as simple care. However, it becomes pivotal in production-scale deployments, which necessitate detection of vital ambient sign. Hence, it proliferates among multidisciplinary research communities, such as healthcare, artificial intelligence, sensing, communication, cognition, privacy, ethics, psychology, etc. All research endeavors implicate personalized interfaces and services, which are compulsory to overcome engagement impairments of aging generations.

Enabling engagement information as a utility promises to improve urban life. Since, knowledge comprises of data from heterogeneous sources, the exploration of inherent economic value is very important. However, the increasing demand of personalized services is demanding rejuvenation in different steps of service-value chain, such as infrastructure-development, content production, delivery. Especially this emphasizes monitoring of improvement on satisfaction of citizen-end, after infrastructure is implicated by operators. However, conventional participation management procedure is intricate and subjective, which involves labor, human and thereby financial expenditure. On the other hand, citizens are ready to pay only for improved network experience, if they are happy about the service, its contexts and conveniences. Therefore, rejuvenation is aim at activity monitoring service-consumption, graceful degradation of performance from the optimal point and consequent quantification option for citizens. In this context, migration of statistical, optimization tools and even multidimensional sciences, such as network-economic, pricing, is essential for user-centric evaluation and control.

Up-to-the-minute urban solutions[4]-[8] deal with engagement constraints, such as, citizen's introvert or uncongenial mind, cognitive impairment, unusual interrelated cerebral or somatic symptoms, latent activity patterns and demand-response between citizen and urban-planner. Hence, participation is monitored through digital or augmented media, such as, social-network, cellular traffic, wearable sensors and connected home-appliances. Engagement is encouraged with knowledge-exchange, societal-mingle or conjugation and situation-awareness However, conventional practices deficit sufficient realization of personal traits, such as monochromatic-choice, disparate impairment, generalization and thematic-representation and rich annotation

Therefore, instinctive instrumentation is envisioned, in which engagement is stimulated with accurate quantification of contexts[9]. Hence, key prerequisites are speckled, which automation should clinch. Primarily, a discovery model is essential, which accurately annotates latent theme from plethora of elderly information. However, the latter stage demands scalability, which illuminates transformation over time and attachment on connected themes. This provides both ease and convenience for remote caregivers in optimized decision making, so that elderly enjoy independent daily- living.

To encounter above confrontations, parametric Bayesian scheme is envisioned, which predicts an autonomous model for uncovering contexts and scaling up with time. Medium-scale measurements are conducted to gather polychromatic experiences of participants. However, an impractical implication of statistical information, may lead to conclusion, that is not apotheosis. Therefore, our developed platform attempts to grapple with some completely baffling ingredients.

A synopsis of paper is as follows, (a) engagement-challenge of urban, ambient augmented thing (section II), (b) motivation of context-aware rejuvenation (section III), (c) practical testimony of procedural development with prototype and assessment (section IV), (d) concluding remark (section V).

# ENGAGEMENT CHALLENGES ON URBAN AMBIENT AUGMENTED THINGS

Global demand of 'urban revamp, healthier and happier generation' requires sustainable plans and executions by policy makers, technologists and public/public organizations. Moreover, social-economic demand of healthier generations acquire recommended practices, such as, ambient assisted living, electronic health, mobile health, and smart health. Last but not the least, with recent proliferation of immersive ecosystems, augmentation is a urban companion, rather than technological hoax. Eventually, paramount challenges are imposed for transfixing citizen with lively stimuli on conventional urban events.

**Citizen Behavior**

Monotonous daily schedule keeps citizen isolated from societal mingle. However, information-exchange is scientifically proven to invigorate communal credence. Hence, minimal but useful information is to be shared in overcoming hustle of excessive data. Trail conduction with subjects, who are reluctant in social engagement, is one of viable solutions. Because, peers evaluate recommendation from posts, comments, pictures and voice-messages by a self-comparison process. Thus, viewing or responding to recommendation drive towards healthier social life.

**Service Visualization**

Visualization is essential in assessing meaningful roles and needs of different user-groups and service-providers. Therefore, context or presence is frequently inferred from ambient signs. Such kind of visualization becomes crucial in adapting severe health or environmental hazards. In this context, spatial and temporal data is authenticated and visualized to a mode, which eases emergency decision making. Personalization Urban augmented services involve personalization without compromising social isolation. However, personalization is harmonious with situation awareness, such as emotion, environmental information or even domestic happening. Therefore, a robust automated mechanism is envisioned, which adapts individual context in scalable manner with smarter devices and appliances.

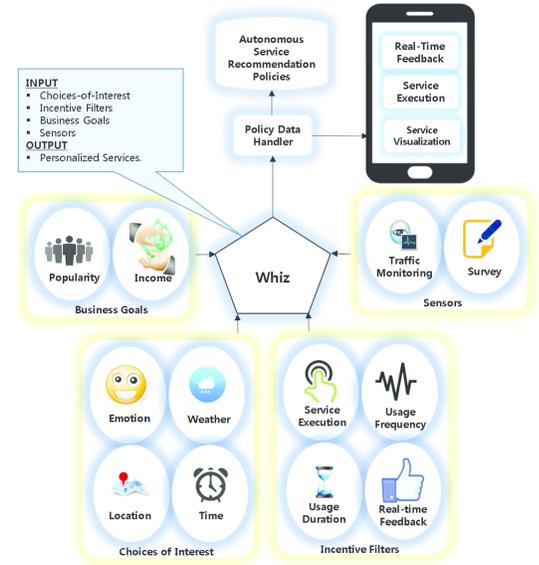

Fig.1: Ambient Augmented Things on Urban Space

**Information Management**

Multi-modal information management is derived from multifaceted public demand. Statistical learning is recognized for superposing manual or automatic labels on sensory data. For example, classification or clustering discovers contextual and temporal features from IoT-appliances.

**Information Fusion**

Personalization is difficult for the involvement of unusual cerebral and somatic symptoms. Hence, it is required to sense and finally fuse information on activity or even presence. For example, sensor readings and crowd intelligence are utilized together in localizing healthcare appliances in indoor environment.

In all cases, hysteria of entertainment requires contextualization of urban activity in supporting ambient experience. However, geographic dispersion and entailed complex technologies are major barriers in this case. Because, mass participation is essential in demonstrating the viability of real-scale deployment. Thus, planners are struggling for an optimal solution, which aggregates small and medium-scale assessment into a feasible outcome.

# SMART CONTEXT AWARE REJUVENATION

The hype of connected Big Data enriches digital humanity, in which augmented presence of smart-objects is sensed from Internet-of-Things(Fig. 1). Eventually, urban planners are encouraged about quantitative outcome of relevant penetration. However, it requires scientific knowledge from multiple disciplines to grapple a holistic view. Therefore, a proper interplay between context, mobility, social-presence is essential for a viable solution. Such contextual resonance demands blending data analytic with knowledge-base to correlate usage-information and context. However, scientific instrumentation conducts proper navigation, which finds similarities or miss-matches by occasional highlight or narrow-down the scope. Therefore, discovery of meaningful pattern and succeeding stochastic adaption convey contextual rejuvenation.

**Big-Data Hype, Digital Community and Monetization**
A buzz is going in the industry about Big Data, augmented presence and relevant quantitative outcome. This actually is originated from the promise of billions of connected devices and social-data Eventually, it is having a great influence on content-generation, delivery and value-oriented service chain. However, the actual influence is realized from the cultural diversity and similarity across generations. Eventually, there is a demand on the market for digital humanity. Such digital presence is inherent in contextual information, rather than outward expressions. Therefore, the generation of meaningful contextual interpretation from content corpora is emerging. Digital presence is merely dependent on participation preference, rather than content-provider objective. However, digital usage patterns pose indirect influence on business-chain. Thus, a bridge is mandatory for mitigating the gap among digital objects, consumer perceptions and business-goals.

**Multidisciplinary Scientific Knowledge**
Rigorous analysis on participation information demands multifaceted knowledge. Therefore, migration of tenants from statistic, economic, psychology and information theory is a common practice. As such, instrumentation grapples scientific information in justified manner. Moreover, systematic methodology leverages information of large user-groups. Data-sets of multiple domains are explored, such as traffic measurement, social-science and psychology. Therefore the size of knowledge-base ranges from tiny access to massive oracle databases. However, amongst varieties solutions, an optimal technique is selected, which quantifies necessary information.

**Interplay between consumer-contexts and revenue**
However, proper interplay between usage, mobility and social-presence is required for commercial citizen services. This is also essential for industry to come closer with citizens by innovative user-centric services. However, service providers sort out potential groups from consumer-usage. Because, they are only curious about financial output before stepping into systematic implementation. On the contrast, it is obviously difficult to figure out contexts without any prior knowledge. Therefore, a feasible recommendation is essential, which speaks for target consumer-groups.

**Qualitative and quantitative assessments of cultural diversity**
In the search for substantial usage patterns, service providers look for affinities between contexts and cultural barriers. Because, consumer behavior is correlated with cultural barriers, which necessitate implication of social-science. Moreover, as digitized information is increasing, it is often cumbersome to ascertain target customer-groups. Conventional methodologies classify consumer-preferences from subjective experiences. However, quantitative interpretation, alongside with qualitative input ease assessment through real-world data validation. Therefore, an automated methodology is envisioned to map information into meaningful consumer-groups, which appear more frequently than others, amongst whom are being observed.

**Context-Aware Rejuvenation Representation**
Recommendation schemes classify contexts with labeled information. However, linguistic expression is scientifically proven different than implicit meaning. Hence, an autonomous instrumentation is required for deriving inner meaning from outward representation. In this context, navigation is useful to link up meaningful data with supplementary one, especially for finding similarities or miss-matches. Moreover, the problem becomes intrinsic due to required interactions of tagged information. A common criteria highlights on contexts by occasionally narrowing down the scope. Because, context is evolved over time and connections among them persist in upcoming iterations. Therefore, uncovering latent contexts and then looking for stochastic meaning, especially, how content is generated, is a desired criteria.

## PROBLEM ASSESSMENT AND PROTOTYPE DEVELOPMENT

Our aim is to develop instinctive instrumentation, where citizen engagement is rejuvenated with accurate quantification of contexts. Hence, key prerequisites are speckled, which automation should clinch. To encounter identified confrontations, our parametric solution predicts an autonomous model for uncovering contexts and scaling up engagement. An impractical implication of statistical information leads to conclusion, that is not apotheosis. Therefore, we attempt to grapple with some completely baffling ingredients through assessments and systematic prototype development. Primarily, a discovery model is sensed, which accurately annotates contexts from plethora of participatory information. However, unwrapping common contexts from plethora of geographically dispersed information is a difficult task. Conventional schemes involve manual configuration by service providers, which is both time and money consuming. Moreover, representation of knowledge from latent contexts, is cumbersome. The instrumentation requires quantification, whether information is matched to that contexts, which scale up over time and are tied within each others.

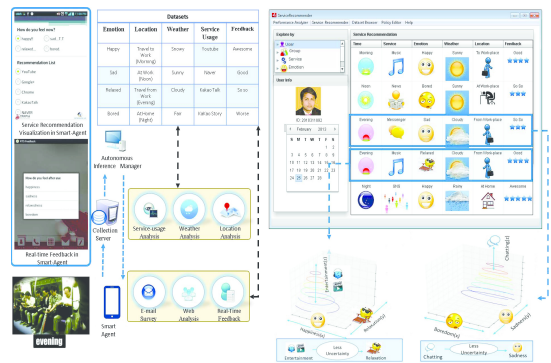

Fig. 2: Urban-centric Ambient Augmented Things Management scenario

**Problem Assessment**

In this context, survey is conducted among 77 participants about their subjective experiences in different emotions, locations, weather and time, etc(Fig.2)[9]. Questionnaires contemplate their perceptions in different contexts (Q1-Q5) or even fondness with Smart-devices (Q6- Q7), media content (e.g. Radio/TV) (Q8-10), social network (Q11-Q13). Opinions are kept as simple as possible, such that participants never turn confused. Hence,several brainstorming sessions are undergone to establish a rhythm in the survey lifespan. Moreover, a short survey time is maintained to engage complete concentration of participants. Because, participants are often at discomfortable zone in paying concentration to phone-call, instant messaging or request for an urgent work. Therefore, their feedback are gathered in leisure periods, while they are having idle time or just finished up with midterm exams, etc. Participants are asked experiences separately, while they are traveling to/from workplace and at workplace. Because, it is scientifically proven that, participants are reluctance in remembrance of consecutive recent events. Absolute ratings with literary descriptions are used, since similar experiences are interpreted by individuals in different ways. Furthermore, psychological conditions have impact on personalized ratings. Survey questions infer emotional status and find its miss-match with service-usage perception. Because, user-ratings are divergent, even in the cases, when their perceptions are similar. Last but not the least, divergent data is collected from several participant-groups, since opinions from a small group of people give abstract view or biased opinion. Then, Smart-device traces are collected from lab people during their daily activities. Information is gathered at the beginning of each online session. Since experiment is conducted on closed door environment, outside weather or static location has little impact on it. However, participants' verbal opinion on 'location/weather'-oriented preference becomes an alternative, but handy option . It is very difficult to capture emotion through Smart-device traffic. Hence survey and traffic information are quantified together to correlate psychological status with usage. Qualitative experience is contingent on bizarre, incautious instincts of participants. Hence, it is validated against objective counterpart to adjust unusual changes. However, its archetypal outcome is rendered through communication pattern of individual or groups. Combination of qualitative and quantitative outcome is worthwhile to standardize methodologies for real-life comparative analysis. A Simple app is presented before participants for feedback-collection. Questionnaires appear during and just after app-usage. User-interface is kept so convenient, that there is little chance of negative impression. Participants' satisfaction and time variant demands are seldom reflected in limited space of qualitative and quantitative assessments. Hence, accumulation of real-time feedback in an online fashion gives the stimuli a new dimension. Generally, engagement and termination stem from network-centric (e.g. stalling), device-specific (e.g. content resolution), content-based(e.g. immersiveness) or even psychological factors . However, substantial reaction is stimulated, when participant is given the thrill of virtual presence in digitized world. The experience is correlated with emotional, cultural or even psychological attachments. Since it is, more or less, fluctuated by profile and tolerance levels, generality promises to add quality at global scale. Furthermore, participants are asked to rate two stimuli in a convenient group discussion environment. Thus, dichotomous choice replaces aforementioned multi-scaled rating and eases decision making. Moreover, it is scientifically proven that, cognitive distance between 'So So' , 'Worst' is different than that of between 'Awesome', Good'. Interactive dialogues overcome such arithmetic obstacles, in addition to perfunctory hurdles, such as unconsciousness or even reluctance.

**Bayesian Engagement Estimation**
A hypothetical conjecture of context-aware skeleton is visualized by dissecting participation information. Annotated contexts are unveiled from citizen database at inaugural phase. Complex association among contexts, which evolve over time, are to be dealt at latter stage. Therefore, uncovering contexts and then scaling up over time speculate a Bayesian model, which is amenable to statistical inference. Participation knowledge-base is ameliorated amidst a generative process, which involves citizen contexts. Actually, the generative process is a joint probabilistic distribution over observed information and latent random variables. Eventually, the conditional distribution of the latent variables, given the knowledge-base turns to be a posterior distribution. The posterior distribution of original problem for finding latent variable from observed information, is expressed as follows Ultimately, the posterior is sum of joint distribution over all possible instances of latent contexts. Owing to this statistical facet, the numerator represents joint distribution of all random variables for any latent variable. Moreover, the denominator comprises of marginal probability distribution of latent variables, which resemble probability of corpora in devised scheme. However, the posterior is not computable due to the denominator, which is exponential large. Hence, engagement estimation is amenable to statistical inference, which approximates suitable probabilistic model using citizen database.

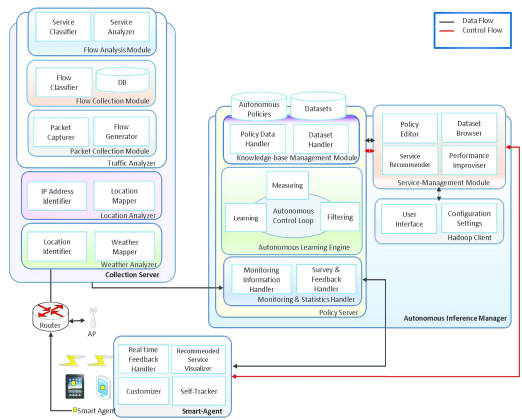

Fig. 3: Prototype architecture

**Systematic Prototype Instrumentation**
Systematic prototype (Fig. 2)[9] is developed in accordance with statistical inference and assessments. It is capable of gathering participants contexts (e.g. weather, location) and recommends personalized service. It includes three major components, namely, smart-agent, collection server and autonomous inference manager. Participant experiences recommended service and gives real-time feedback on smart-agent. On the other hand, collaboration server separates location, time and weather information by analyzing traffic. Furthermore, autonomous inference manager comprises of policy and service management modules. Policy management is concerned with autonomous service, data-set and recommendation management. On the other hand, service management is concerned with policy configuration, service or traffic visualization, etc.

## RELATED WORK

In FriedgeNet[4], nutrient information mounted on standard refrigerator is propagated on cloud-based community. However, particular challenge is imposed on monochromatic choice, such as nutrition-intake. Nocturnal[5] regulates behavior and circadian rhythms with connected network of body sensor nodes and smart-appliances. However, special care is to be taken for citizen with handicrafts. Moreover, the need for situation-awareness is realized, which could have amplified lively appreciation on participation. Bridge[6] leverages wireless sensor network, home-appliances and behavioral pattern in delivering augmented resident services. However, generalization of the framework is specially promising for those, whose impairments restrict them from persuading autonomous activities. Sphere[7] manages connected care from healthcare things or objects. However, requirement of robust annotation is realized for adapting practical well-being scenarios. A smart-phone-based platform[8] leverages ubiquitous access on citizen information. Because, some disparate impairments demand unique procedural practice. However, as impairments are related with multiple modalities, best practice is difficult to achieve in home scenarios.

## CONCLUSION

The contemplation of urbanization and elderly generation motivates the exploitation of citizen participation information for personalized services. Fortunately, human mind is originally attached with societal mingle or engagement, which is resembled through inherent desire for company in events, despite of outward solitude-expression. However, recommended practices deficit proper realization of personal traits, such as monochromatic or disparate preference, automated mobility and robust annotation. Improper assessments on objectives of various user-groups bring unfavorable impact on engagement, rather than warmth acceptability. In this circumstance, digital and augmented presence of smart things and objects paves the way for personalized services. Eventually, context awareness is realized as possibilities for the overall acceptability of groundbreaking citizen services. Ours is a preliminary step in harnessing the blue ocean, while meeting major pitfalls. In this context, parametric Bayesian scheme resembles complex interactions among contexts, while scaling up over time. Consequently, prototype implementation is demonstrated for the accurate rejuvenation of engagement by quantifying participants' opinions, usages or even feed-backs.

| Project | Agenda | Engagement obstacles | Reju-venation issues |
|---|---|---|---|
| FriedgeNet [4] | Urban-centric communal credence by dietary information-exchange. | Citizen psychology | Information exchange |
| Nocturnal [5] | Citizen service visualization from ambient signs. | Service Visualization | Public demand analysis |
| Bridge [6] | Urban resident service with augmented experience. | Persona-lization | Psychology assessment |
| Sphere [7] | Connected health management from urban things and objects. | Information manage-ment | Context resonance |
| Smart-phone-based platform [8] | Ubiquitous citizen service for disparate impairments. | Information fusion | Activity inference |
| Psychic [9] | Personalized service management of urban connected things. | -Persona-lization. -Service manage-ment | Context resonance |

## REFERENCES

1. A. Al-Fuqaha, M. Guizani, M. Mohammadi, M. Aledhari and M. Ayyash, "Internet of Things: A Survey on Enabling Technologies, Protocols, and Applications," in IEEE Communications Surveys & Tutorials, vol. 17, no. 4, pp. 2347-2376, Fourthquarter 2015.
2. Cook, Diane, and Sajal Das. Smart environments: Technology, protocols and applications. Vol. 43. John Wiley & Sons, 2004